\documentclass[aps, pra, preprint, groupedaddress, floatfix, longbibliography]{revtex4-2} 
\usepackage{graphicx} 
\usepackage{amsmath} 
\usepackage{amssymb} 
\usepackage{amsfonts} 
\usepackage[english]{babel}
\usepackage{appendix}
\usepackage[colorlinks=true]{hyperref}
\usepackage{url} 
\usepackage[utf8]{inputenc} 
\usepackage[T1]{fontenc} 

\hypersetup{
 colorlinks=true,
 linkcolor=blue,
 filecolor=magenta,
 urlcolor=cyan,
 citecolor=blue,
}

\newcommand{\de}{Discovery Engine}
\newcommand{\cnm}{Conceptual Nexus Model}

\begin{document}

\title{The \de{}: A Framework for AI-Driven Synthesis and Navigation of Scientific Knowledge Landscapes}


\author{Vladimir Baulin}
\email{vbaulin@activeinference.institute}
\affiliation{Active Inference Institute, Crescent City, California, 95531, USA} 
\affiliation{Universitat Rovira i Virgili, Tarragona, Spain} 

\author{Austin Cook}
\affiliation{Active Inference Institute, Crescent City, California, 95531, USA} 

\author{Daniel Friedman}
\affiliation{Active Inference Institute, Crescent City, California, 95531, USA} 

\author{Janna Lumiruusu}
\affiliation{Active Inference Institute, Crescent City, California, 95531, USA} 

\author{Andrew Pashea}
\affiliation{Active Inference Institute, Crescent City, California, 95531, USA} 

\author{Shagor Rahman}
\affiliation{Active Inference Institute, Crescent City, California, 95531, USA} 

\author{Benedikt Waldeck}

\affiliation{Active Inference Institute, Crescent City, California, 95531, USA} 

\date{\today}

\begin{abstract}
The prevailing model for disseminating scientific knowledge relies on individual publications dispersed across numerous journals and archives. This \textit{legacy} system is ill suited to the recent exponential proliferation of publications, contributing to insurmountable information overload, issues surrounding reproducibility and retractions. We introduce the \de{}, a framework to address these challenges by transforming an array of disconnected literature into a unified, computationally tractable representation of a scientific domain. Central to our approach is the LLM-driven distillation of publications into structured "knowledge artifacts," instances of a universal conceptual schema, complete with verifiable links to source evidence. These artifacts are then encoded into a high-dimensional Conceptual Tensor. This tensor serves as the primary, compressed representation of the synthesized field, where its labeled modes index scientific components (concepts, methods, parameters, relations) and its entries quantify their interdependencies. The \de{} allows dynamic "unrolling" of this tensor into human-interpretable views, such as explicit knowledge graphs (the CNM graph) or semantic vector spaces, for targeted exploration. Crucially, AI agents operate directly on the graph using abstract mathematical and learned operations to navigate the knowledge landscape, identify non-obvious connections, pinpoint gaps, and assist researchers in generating novel knowledge artifacts (hypotheses, designs). By converting literature into a structured tensor and enabling agent-based interaction with this compact representation, the \de{} offers a new paradigm for AI-augmented scientific inquiry and accelerated discovery.
\end{abstract}

\maketitle 


\section{Introduction}

Scientific progress relies on the effective accumulation, synthesis, and critical evaluation of knowledge. Traditionally, the well-documented, peer reviewed publication served as the primary standard for filtering and disseminating credible findings within the scientific community. Recently, however, we are witnessing an unprecedented acceleration in research output, a veritable explosion of scientific publications across all disciplines \cite{park_papers_2023}. Yet, this very abundance creates a paradox: the sheer volume threatens to overwhelm the mechanisms designed for its assimilation and synthesis. Researchers, even within highly specialized subfields, face an almost insurmountable challenge in keeping abreast of relevant developments, integrating disparate findings, and identifying the truly novel signals amidst the noise \cite{baker_1500_2016}. This information overload contributes to disciplinary fragmentation, hindering the cross-pollination of ideas essential for disruptive innovation \cite{rong_40_2025}. Furthermore, persistent concerns regarding "reproducibility crisis" \cite{baker_1500_2016}, predatory journals, inflation of research areas\cite{kang_limited_2025}, growing retractions and the potential influences of bibliometrics on research direction \cite{tyson_public_2024} highlight systemic challenges in validating and prioritizing scientific contributions to fundamental knowledge.


We argue that a significant contributing factor to these challenges lies in the \textbf{persistent reliance on a document-centric model} of scientific communication, a legacy format optimized for human reading but ill-suited for large-scale machine search, analysis and synthesis. A scientific paper, typically disseminated as a PDF or typeset documents, presents findings embedded within a linear narrative. This narrative necessarily intertwines background research, core factual claims, detailed methodological parameters, quantitative results, and underlying theoretical assumptions with authorial interpretation, contextual framing, and rhetorical choices \cite{mongillo_synaptic_2024}. While vital for conveying the scientific journey, this format makes the automatic extraction and integration of the underlying, reusable knowledge components, \textit{e.g.} the precise definitions, the specific experimental conditions, the verifiable parameter values, the explicitly tested relationships, an arduous, often manual, and inherently lossy process.

Extracting the verifiable, relational "atomic facts" or knowledge components from millions of such documents remains an intractable challenge for purely automated systems without significant human intervention or the adoption of more structured representations. Key relationships, parameter dependencies, methodological nuances, and implicit assumptions often remain locked within the prose, lacking the standardization, granularity, and explicit relational structure required for effective machine processing and deep synthesis across the corpus.

To address this bottleneck and unlock the next level of scientific progress, particularly in an era increasingly shaped by artificial intelligence (AI), we need to transition towards a knowledge infrastructure that is fundamentally machine-readable and designed for automatic extraction and synthesis into new narratives. This requires embracing principles like FAIR (Findable, Accessible, Interoperable, Reusable) \cite{wilkinson_fair_2016,jacobsen_fair_2020}, ensuring that the core components of scientific knowledge are structured not just for human consumption, but for computational analysis and integration by intelligent agents \cite{wooldridge_1_1996}. We need to move beyond viewing the literature as a digital library of static documents towards conceptualizing it as a dynamic, interconnected graph of verifiable knowledge components.

Here we introduce the \de{} (DE) as a methodology and conceptual platform designed to architect this transition. The DE moves away from the analysis of isolated publications towards the automated synthesis of entire scientific fields into dynamic, structured knowledge repositories. The central construct generated and maintained by the DE is the \textbf{\cnm{} (CNM)}. This CNM is envisioned as an evolving, AI-curated knowledge graph that serves as a cognitive "World Model" for a specific research domain \cite{tsividis_human-level_2021}, capturing not just individual facts in human language but the intricate web of relationships between concepts, methods, findings, and theories. The \de{} is not a static model, but an interactive ecosystem incorporating AI agents designed to assist researchers in navigating, analyzing, and extending this synthesized knowledge landscape.

It is noteworthy, that LLMs are employed not for unconstrained generation, but as tools for \emph{Structured Knowledge Distillation}, guided by adaptive, formally-defined templates (the structured and self-consistent schema for processing scientific documents, specific for each field and dynamically adapting). This process extracts granular, verifiable knowledge components from source literature, ensuring consistency and traceability. These extracted components are then represented using formalisms designed to capture both semantic meaning and structural relationships. This might involve hybrid approaches combining dense vector embeddings for semantic similarity \cite{piantadosi_why_2024, kriegeskorte_representational_2008} with techniques capable of encoding explicit compositionality and relational structure, potentially drawing inspiration from Vector Symbolic Architectures (VSA) \cite{kleyko_vector_2022, jr_computational_2018} or the abstract relational language of Category Theory (CT) \cite{phillips_what_2022, yuan_categorical_2023, pan_token_2024}.

The resulting machine-readable CNM provides the essential substrate for a new generation of AI-driven scientific tools that are optimized for machines, while also being useful for computational multimodal generation and human accessibility. AI agents operating on this structured knowledge graph can perform tasks far beyond simple information retrieval \cite{wooldridge_1_1996}. They can navigate complex conceptual landscapes, identify deep structural analogies between different research areas \cite{lu_emergence_2019, friedman_fieldshift-2_2024}, detect subtle inconsistencies or contradictions across studies \cite{sejdinovic_hypothesis_2012}, systematically pinpoint under-explored regions or knowledge gaps \cite{buehler_agentic_2025-1}, and assist researchers in formulating novel, data-grounded hypotheses \cite{obrien_machine_2024}. The DE, therefore, aims to foster a synergistic relationship between human researchers and AI systems, where AI handles the large-scale synthesis and structural analysis, freeing human intellect to focus on interpretation, creativity, and critical evaluation.

\section{The Conceptual Nexus Model: A Structured, Computable Representation of Scientific Knowledge}
\label{sec:cnm}

One of the aims of the \de{} methodology is the creation and update of the \cnm{} (CNM). It serves as a dynamic, structured representation of a scientific field, moving beyond the limitations of traditional, document-based knowledge storage. Its construction integrates AI-driven extraction processes with formal representation schemas, enabling the transformation of textual knowledge into a computable and interconnected network designed for both human exploration and automatic AI-based analysis. The aspiration is to construct a CNM of such fidelity and completeness that it begins to mirror the intrinsic logical structure of the scientific domain itself. In such a scenario, analogous to how physical systems compute their own evolution according to underlying natural laws \cite{lloyd_ultimate_2000}, an AI agent could interact with this validated "World Model" in a zero-shot fashion, inferring new knowledge or validating hypotheses through computational exploration of the CNM's structure without requiring specific human feedback for each instance. The observation that fundamental physical laws are often expressible in remarkably compact mathematical forms \cite{wigner_unreasonable_1960} suggests that the parameter space of core scientific principles within a domain might be learnable and navigable by AI, provided a sufficiently structured and verified knowledge representation like the CNM. 

\begin{figure*}[ht]
\includegraphics[scale=0.22]{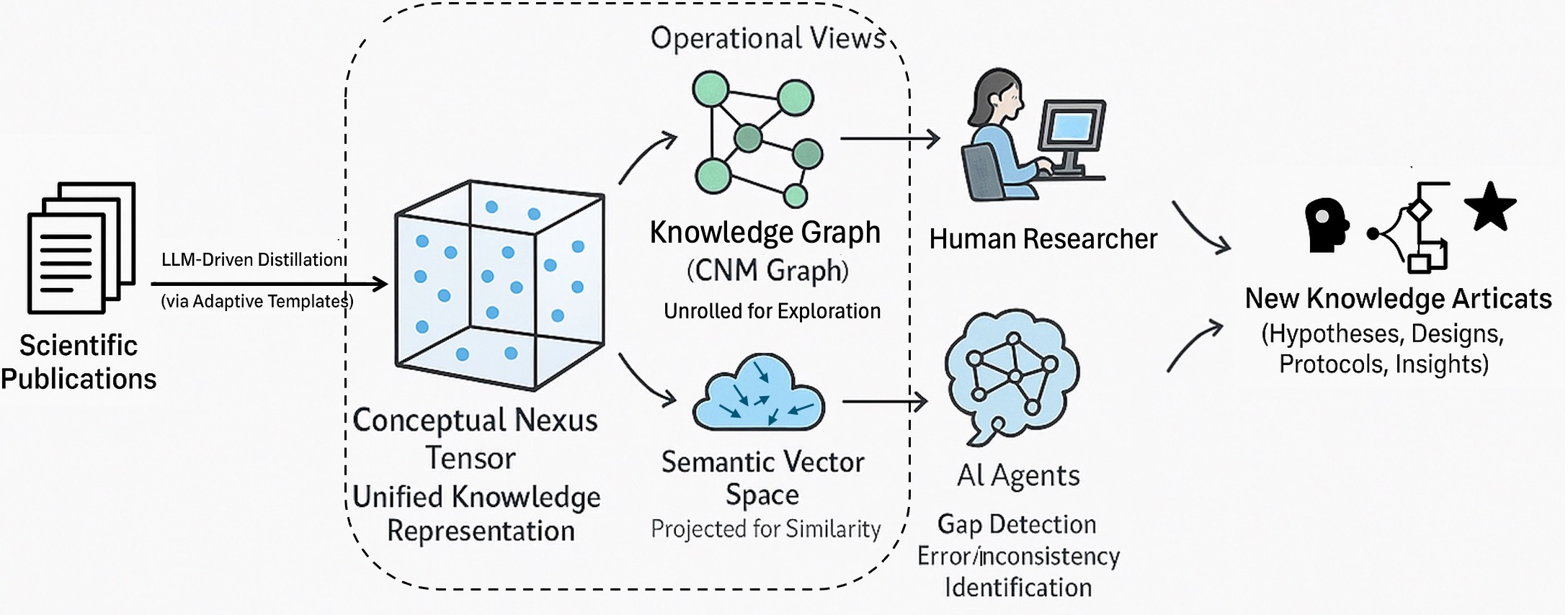}
\caption{Conceptual Nexus Model for distillation of the knowledge into machine-readable format ready for human and agent exploration and machine-facilitated discoveries. \label{fig:schematic}}
\end{figure*}

A fundamental challenge in synthesizing scientific knowledge lies in transforming the heterogeneous, often narrative-driven content of publications into such a structured, computable format, Fig. \ref{fig:schematic}. The traditional publication format, while effective for detailed human communication, inherently embeds core findings, methods, parameters, and conceptual relationships within natural language, making large-scale, automated extraction and comparison extremely difficult \cite{mongillo_synaptic_2024}. However, one of the requirements for reproducibility of results, is that the logical structure of experiments must be articulated with precision to allow for replication, a process that requires unambiguous language or unified schema in the description of steps \cite{buzbas_logical_2023}. 

\begin{figure}[ht]
\centering
 
 \includegraphics[scale=0.5]{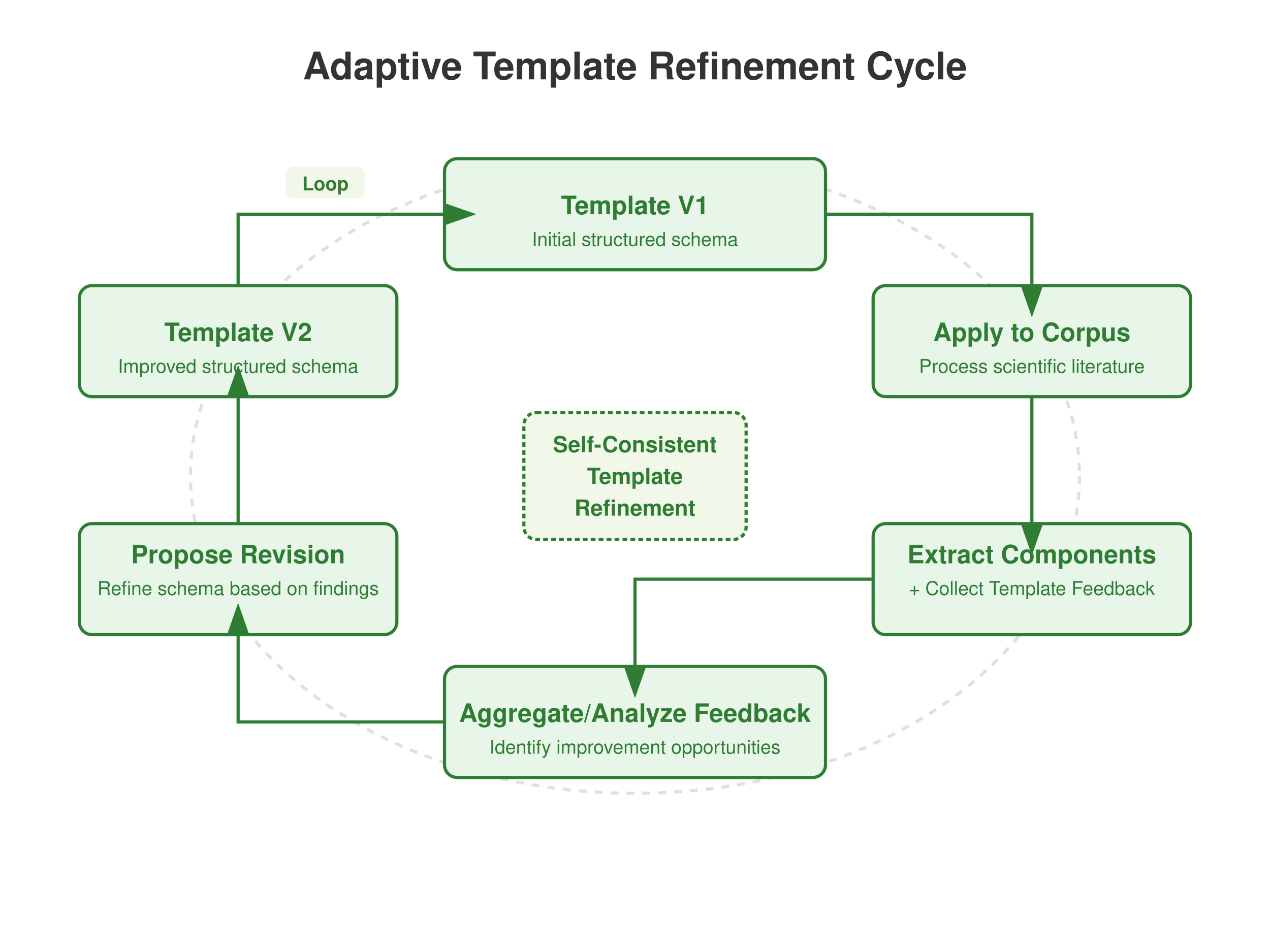}
 
 \caption{The self-consistent template refinement cycle in the DE until the template and the corps of literature become consistent.} \label{fig:template_evolution}
\end{figure}

Instead of utilizing traditional bibliometrics as measures of the credibility of a publication, DE implements a robust yet flexible FAIR-aligned Verifiability and Robustness Scoring system for evaluating each knowledge artifact in the CNM. These scores could evaluate not the perceived importance of a finding, but its inherent scientific utility and trustworthiness in a dynamic and multi-faced way. For example, a "Findability \& Accessibility" score would reflect clearly defined components and well-documented data/methods, "Interoperability" would be gauged by the artifact's ability to connect with and be understood in the context of diverse concepts within the CNM, facilitated by standardized mapping to the Universal Concept Schema, "Reusability" (and overall robustness) is assessed through metrics like: Feasibility (e.g., are methods practically implementable, are parameters within realistic ranges?), Confirmation Strength (e.g., how many independent studies or lines of evidence within the CNM support this artifact or its constituent claims?), Evidence Linkage (e.g., directness and quality of provenance trails to primary data and explicit experimental validation), and Predictive Consistency (e.g., how well does this artifact align with, or get corroborated by, analogous findings or structurally similar systems identified elsewhere in the CNM?). These dynamic and adaptive scores would provide a multi-faceted, evidence-grounded measure of an artifact's potential for reliable integration and reuse in future discovery, rather than a problematic singular and fixed metric of academic influence.

\begin{figure*}[ht]
\includegraphics[scale=0.30]{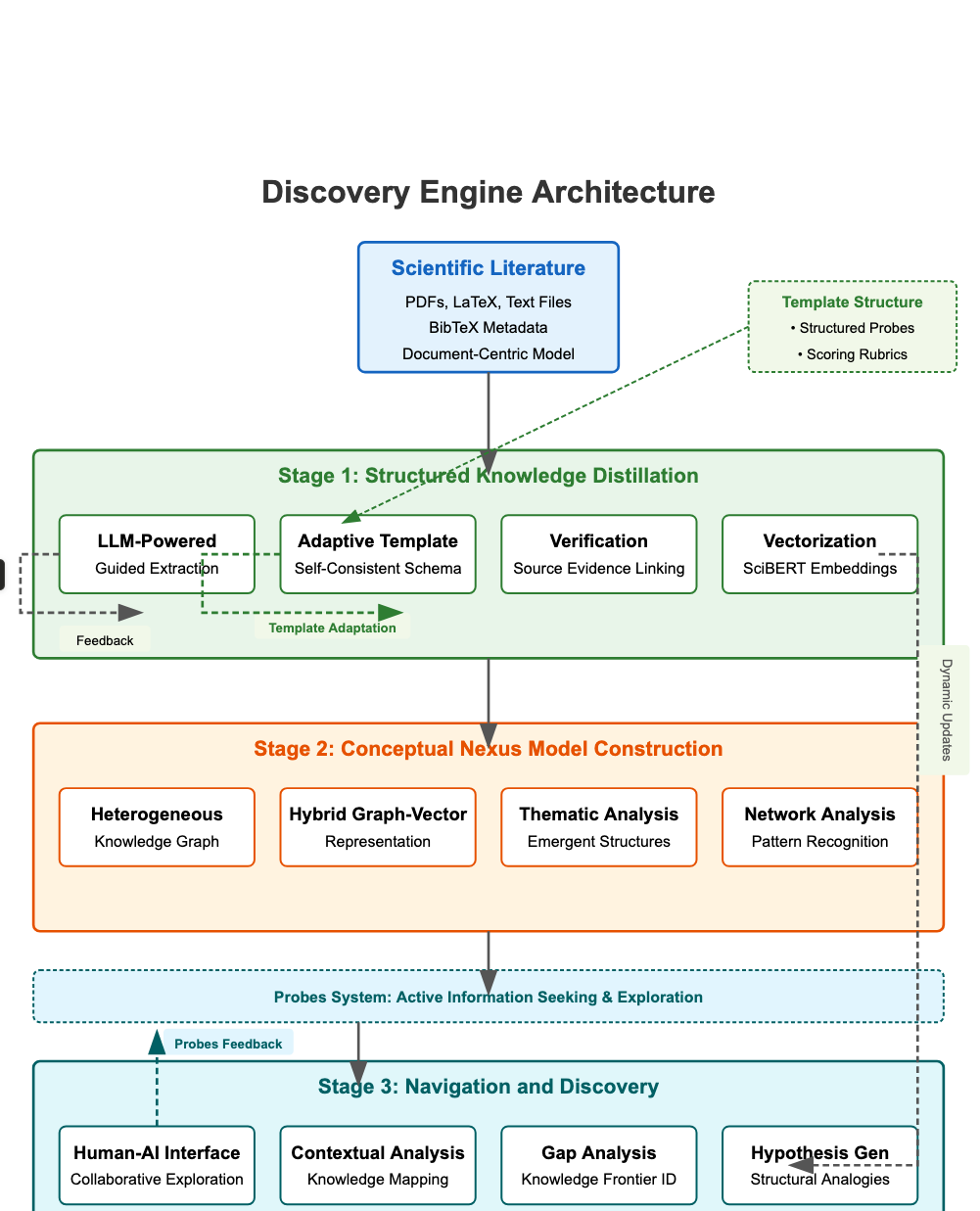}
\caption{Conceptual architecture of the \de{} framework. \label{fig:knowledge_pipeline}}
\end{figure*}

Recognizing the limitations of purely statistical NLP methods for capturing the precise semantics and relationships critical to scientific discourse, this process employs Large Language Models (LLMs) as sophisticated analytical engines, carefully guided by formally specified, adaptive templates (details in Appendix and template schema, Fig. \ref{fig:template}). These templates act as dynamic schemas, defining the types of knowledge components (nodes) and relationships (edges) to be extracted, thereby ensuring consistency and structure across diverse sources. Critically, the process mandates that LLMs provide justifications and link extracted components directly to evidence within the source publication, fostering verifiability and traceability essential for addressing reproducibility concerns \cite{baker_1500_2016}. These templates (drawing conceptual inspiration from Category Theory's focus on structure and mappings \cite{phillips_what_2022, yuan_categorical_2023}, and potentially analyzable using techniques inspired by Graph Isomorphism Networks \cite{buehler_graph-aware_2025}) define the explicit schema for knowledge extraction. LLMs are tasked with instantiating this schema for each publication, identifying specific entities and relationships and providing auditable links back to the source text, thus ensuring verifiability and provenance. Furthermore, the template framework incorporates a \textit{self-consistent refinement loop} (Fig. \ref{fig:template_evolution}), whereby feedback on the template's applicability gathered during corpus processing informs its subsequent evolution \cite{buehler_agentic_2025-1}, ensuring the schema dynamically adapts to the specific nuances and evolving structure of the knowledge within a given field \cite{buehler_self-organizing_2025}.

The conceptual architecture of the \de{} (Fig. \ref{fig:knowledge_pipeline}), outlines the structured knowledge distillation process, the hybrid graph-vector representation forming the CNM, and the methodologies for synthesizing and analyzing the resulting knowledge landscape. This framework enables systematic identification of knowledge gaps and inconsistencies, supports the generation of novel hypotheses grounded in synthesized evidence, and provides a foundation for next level human-AI collaboration in science. We argue that adopting such computable knowledge models is not merely an incremental improvement but a necessary evolutionary step for navigating the complexity of modern science, fostering deeper understanding, enhancing research integrity, and ultimately accelerating the engine of discovery by transforming the way we interact with and build upon our collective knowledge.

The extracted, attributed components populate the CNM, structured as a heterogeneous and hierarchical knowledge graph (Fig. \ref{fig:knowledge_pipeline}). This graph utilizes distinct node types for diverse scientific entities and richly typed edges to represent their specific interrelations. While the graph forms the primary knowledge structure, complementary dense vector representations (Fig. \ref{fig:knowledge_pipeline}) are derived to enable efficient semantic similarity search, machine learning integration, and analogy finding \cite{kriegeskorte_representational_2008, obrien_machine_2024}.

\subsection{Structured Knowledge Distillation via Guided AI and Adaptive Templates}
\label{subsec:principles}

\begin{figure}[tbh] 
\centering
\includegraphics[width=\textwidth]{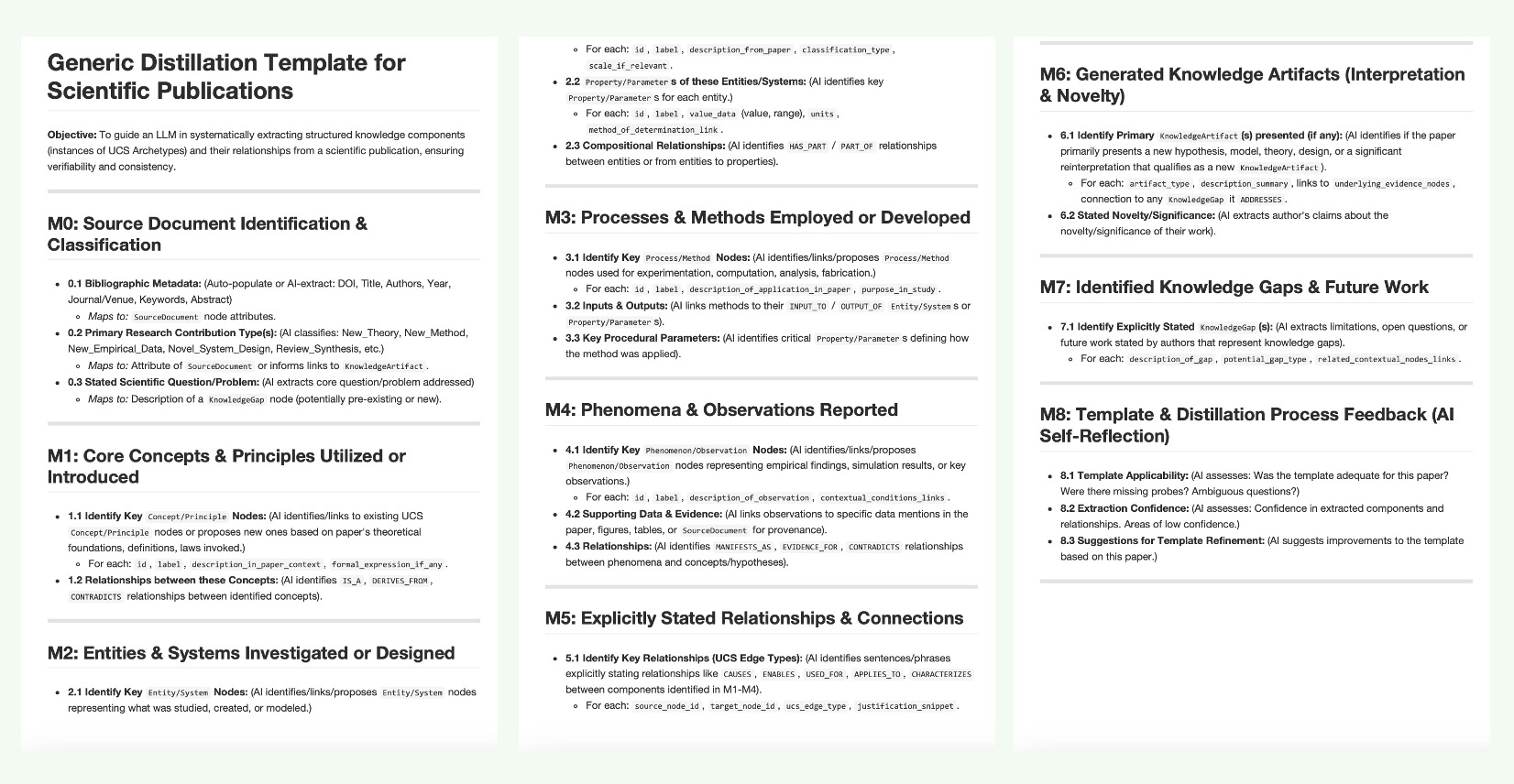}
\caption{Conceptual architecture of the Distillation Template. }
 \label{fig:template}
\end{figure}

The objective is to move beyond simple document summaries or keywords extraction towards isolating granular, verifiable \emph{knowledge components}. These components serve as the \emph{building blocks} for representing the core concepts of a field. What constitutes a "concept" is itself a complex issue debated in philosophy and cognitive science \cite{stich_concepts_2003, piantadosi_why_2024}. Within the DE framework, concepts are operationalized through the structured set of components: nodes and edges (see Appendix \ref{app:schema}) defined by a given appropriate kind of template. The template, see Figure \ref{fig:template}, essentially defines which aspects, such as specific mechanisms, underlying principles, key quantitative parameters, methodological details, reported outcomes, limitations, and explicit relationships to other concepts, collectively constitute the machine-readable representation of a concept or finding within the synthesized \cnm{}. From this standpoint, conceptual meaning arises from both its local document-centric context as well as its structured role and relations as a contribution to a larger knowledge framework \cite{piantadosi_why_2024, fodor_connectionism_1988,kuhn_structure_1994}.

The distillation process is initiated by defining the scope of interest, which can inform the initial structure or focus of the template. This template, implemented as a structured plain text document (e.g., Markdown), serves as a dynamic schema and an analytical lens. It contains a detailed set of probes, ranging from requests for qualitative descriptions and justifications to specific fields for quantitative parameters (requiring values and units) and comparative scores based on predefined rubrics. The modular design (e.g., sections covering aspects like system scope, objectives, implementation, etc.) ensures a systematic interrogation of each publication in a standardized form.

The extraction itself is performed by an LLM. Unlike open-ended generative tasks, the LLM's role here is carefully constrained: it is instructed to act as a rigorous analytical agent, tasked with populating the provided template based \emph{solely} on the content of a single input publication. The prompts emphasize precision, requiring the LLM to extract specific and well-defined values such as parameters and values of experimental techniques, adhere strictly to the template format, and critically, provide direct textual justifications from the source document for every extracted component and assigned score. This focus on verifiable extraction linked to source evidence is paramount for building a trustworthy knowledge base \cite{baker_1500_2016} and mitigating potential LLM hallucination or fabrication. While LLMs have known limitations \cite{rogers_primer_2021}, their ability to follow complex instructions and structured information extraction makes them suitable engines for this distillation task when properly guided and constrained by the template.

An innovative aspect of this methodology is the \textit{self-consistent refinement loop} designed to iteratively adapt the extraction template towards an optimal representation of the scientific domain (Fig. \ref{fig:template_evolution}). This adaptive process draws conceptual parallels with evolutionary approaches in AI-driven discovery, such as AlphaEvolve \cite{novikov_alphaevolve_nodate}, where solutions (in our case, the template schema) are iteratively improved based on evaluative feedback. Here, the "evaluation" is not against a predefined fitness function but against the collective structure of knowledge expressed across the entire processed corpus. Specifically, the template incorporates meta-probes that instruct the LLM to reflect on its own capacity to capture the full scope of information within each processed publication—identifying knowledge components or relational nuances present in the text that the current template iteration fails to adequately represent. This feedback, systematically aggregated across numerous documents, reveals statistical patterns of misfit or recurring representational gaps. Periodically, this aggregated feedback drives AI-assisted (or human-curated) revisions to the template structure, such as adding new descriptive probes, refining existing component definitions, or expanding relationship taxonomies. This iterative cycle \cite{buehler_agentic_2025-1} enables the template schema to converge toward dynamical alignment, not with an initial user guess, but with the inherent structure and conceptual distinctions manifest in the broader scientific literature. The goal is to achieve a template that is comprehensive, minimally ambiguous, and acts as a dynamically reconciled consensus model of how the scientific field itself organizes and articulates its findings, becoming maximally fit for distilling the essence of the corpus \cite{buehler_self-organizing_2025}.

This feedback, collected systematically across the entire corpus of processed publications, provides a rich dataset reflecting the alignment between the current analytical framework (the template) and the actual structure of knowledge presented in the literature. This aggregated feedback is then used to inform periodic revisions of the template itself, a process potentially also assisted by LLMs tasked with synthesizing the feedback and proposing specific modifications (e.g., adding new probes, refining definitions, adjusting scoring rubrics).

This iterative cycle of extracting components is based on  the current template, collecting
feedback on the template’s adequacy, and using aggregated feedback to refine the template
for the next iteration and establishes a self-consistent loop \cite{buehler_agentic_2025-1}. The justification for this approach draws parallels with self-consistent field methods in physics or iterative refinement algorithms in machine learning. The core idea is that a good representation schema (the template) should accurately and consistently capture the phenomena it aims to describe (the knowledge in the literature). If the template consistently fails to capture important aspects or generates ambiguity across many documents, it indicates a mismatch. By using the feedback derived from applying the template to the data (the literature corpus), we iteratively adjust the template itself. This process aims to minimize the "tension" or discrepancy between the template's structure and the structure inherent in the data of the corpus of the publications (consensus). It converges, ideally, towards a state where the template provides a stable, comprehensive, and minimally ambiguous schema for distilling the knowledge within the target corpus, achieving a form of self-consistency between the analytical framework and the knowledge base \cite{buehler_self-organizing_2025}. This dynamic adaptation is vital for fields where concepts and methodologies evolve.

This process of guided distillation and iterative template refinement is hypothesized to effectively "extract" concepts from the literature. By forcing the analysis into a hierarchical structured format defined by the template, and then refining that template based on how well it fits the corpus, the system learns to identify and delineate the key recurring informational units (parameters, mechanisms, methods, relationships) that constitute the conceptual building blocks of the field. The converged template implicitly defines the relevant concepts and their constituent features as understood from the corpus. For example, repeated difficulties in distinguishing two types of control mechanisms using the initial template probes would lead to refined probes that better capture the distinguishing features, effectively solidifying the representation of those distinct concepts within the framework.

The distilled knowledge components, extracted according to the converged template, form the input for the subsequent representation stage. Here, techniques inspired by Vector Symbolic Architectures (VSA) and Hyperdimensional Computing (HDC) \cite{kleyko_vector_2022, jr_computational_2018, frady_computing_2022} offer a promising approach for compressing this structured information into a hierarchical vector database.
This VSA/HDC approach offers several advantages relevant to the DE's goals: (1) \emph{Compression}: Complex structured information is compressed into fixed-size vectors. (2) \emph{Robustness}: High-dimensional vectors are inherently robust to noise and partial information, reflecting the often incomplete or noisy nature of scientific data extraction \cite{kleyko_vector_2022}. (3) \emph{Computation}: Vector operations (addition, multiplication, permutation, similarity calculation via dot product/cosine/Hamming distance) provide a basis for querying, comparing, and reasoning about the structured knowledge directly in the vector space \cite{frady_computing_2022}. A concrete example is the challenge of representing recurring methodological descriptions. Thousands of papers might describe variations of a standard protocol (e.g., PCR, Western Blot, specific simulation setup). Instead of storing redundant or incomplete textual descriptions, the DE uses the template to extract key parameters and variables. VSA/HDC could then represent the core protocol as a base vector, and specific variations (e.g., different annealing temperatures, antibody concentrations, simulation time steps) could be encoded by binding parameter-value vectors to the base vector or modifying specific components. This creates a highly compressed, structured representation where similarities and differences between protocols can be computed via vector operations, automatically identifying clusters of similar methods or specific deviations.

This entire process relies on several key assumptions and faces potential challenges:
\begin{itemize}
 \item \textbf{LLM Fidelity Assumption:} The process assumes that the guided LLM can accurately extract the specified components and provide faithful justifications without significant hallucination or misinterpretation, especially concerning quantitative data and nuanced relationships. Ongoing LLM development and rigorous validation are crucial \cite{lu_fine-tuning_2025}.
 \item \textbf{Template Expressiveness Assumption:} It assumes that a structured template, even an evolving one, can adequately capture the richness and complexity of scientific concepts and arguments, which often involve subtle nuances, implicit assumptions, and complex logical structures potentially challenging to fit into predefined probes \cite{stich_concepts_2003}.
 \item \textbf{Convergence Assumption:} The iterative template refinement process is assumed to converge towards a stable and useful state. However, convergence is not guaranteed. Oscillations, divergence, or convergence to a suboptimal local minimum are possible, depending on the quality of feedback, the nature of the literature, and the template update mechanism. Continuous monitoring and potential human oversight are likely necessary.
 \item \textbf{Bias Amplification Risk:} Biases present in the literature corpus or the initial template design could potentially be amplified through the iterative process if not carefully managed. The feedback mechanism needs to be robust against reinforcing systematic omissions or misrepresentations.
 \item \textbf{Scalability:} Processing millions of papers and managing the feedback loop for template evolution presents significant computational and logistical challenges, requiring efficient algorithms and infrastructure.
\end{itemize}
These challenges represents the validity limitations of this stage. For instance, if an LLM consistently fails to extract specific quantitative data accurately despite template refinements, or if the template structure fundamentally cannot represent a newly emerging paradigm in the field, the distillation process will be compromised, impacting the quality of the resulting CNM. Similarly, if the iterative refinement loop fails to stabilize or converges to a poorly representative template, the self-consistency goal is not met.

Despite these challenges, the proposed methodology of guided AI distillation coupled with adaptive template refinement offers a principled and potentially powerful approach for converting the vast scientific literature into a structured, computable knowledge base, paving the way for the synthesis and discovery stages of the \de{}. The structured output adheres inherently to FAIR principles: (i) being findable (via metadata), (ii) accessible (machine-readable format), (iii) interoperable (through the consistent schema), and (iv) reusable (as granular components). It serves as the verified, componentized input for subsequent representation within the pipline.

\subsection{Unified Knowledge Representation: The Conceptual Nexus Tensor}
\label{subsec:tensor_representation}

Following structured distillation (Sec. \ref{subsec:principles}), the heterogeneous collection of extracted knowledge artifacts---each a rich instantiation of the Universal Concept Schema (UCS, Appendix \ref{app:schema}) comprising textual descriptions, quantitative parameters, symbolic types, and explicit relational links—is encoded into a unified, high-dimensional mathematical object: the \textbf{Conceptual Nexus Tensor ($T_{\text{CNM}}$)}. This tensor serves as the primary, computationally tractable representation of the synthesized scientific knowledge field within the \de{}. It moves beyond a simple collection of linked data points to embody a latent space capturing the multifaceted interdependencies of scientific knowledge.

The construction of $T_{\text{CNM}}$ is a critical encoding step. Its modes (axes) are rigorously defined by the UCS archetypes and key metadata dimensions extracted from the source literature. For instance, distinct modes could index:
\begin{itemize}
    \item \texttt{UCS Node Archetypes} (e.g., `Concept/Principle`, `Process/Method`, specific instances like `natural-selection` or `PCR`).
    \item \texttt{UCS Relationship Archetypes} (e.g., `CAUSES`, `USES\_INPUT`, `EVIDENCE\_FOR`).
    \item \texttt{Contextual Dimensions} derived from knowledge artifact attributes (e.g., discretized parameter values, experimental conditions, temporal markers, semantic features from textual descriptions via pre-trained embeddings \cite{piantadosi_why_2024, kriegeskorte_representational_2008}.
    \item \texttt{Provenance Indicators} (e.g., clusters of source publications, research sub-domains).
\end{itemize}
An entry $T_{i,j,k,...}$ in this tensor would quantify the existence, strength, probability, or information-theoretic measure (e.g., mutual information) of the relationship between the elements indexed by $i, j, k, ...$ across their respective modes. For example, $T_{\text{ConceptA, ProcessB, CAUSES}}$ could store a value indicating the strength of evidence that Concept A causes Process B. This explicit indexing transforms the initial array of disconnected publications into a structured and compact tensor that inherently models n-ary relationships and interdependencies. Methods for populating such a tensor could range from direct encoding of extracted structured relations to learning procedures that map graph structures and component features into this tensorial space, potentially using techniques inspired by tensor factorization for knowledge graph completion \cite{balazevic_tucker_2019} or graph neural networks that learn to populate tensor entries \cite{ren_neural_2023}.

While $T_{\text{CNM}}$ itself is a dense, high-dimensional object primarily intended for machine operation, the \de{} provides mechanisms to "unroll" or project it into human-interpretable views:
\begin{itemize}
    \item \textbf{Knowledge Graph View (The CNM Graph):} A primary projection renders $T_{\text{CNM}}$ as the explicit, heterogeneous Conceptual Nexus Model graph discussed previously (Fig. \ref{fig:knowledge_pipeline}). Nodes correspond to UCS artifact instances (derived from tensor mode indices or specific tensor fibers), and edges are instantiated based on significant tensor entries representing explicit relationships. This view supports human navigation, qualitative exploration, and symbolic graph algorithms.
    \item \textbf{Semantic Vector Space Views:} Other projections can generate task-specific vector embeddings for concepts or artifacts by, for example, slicing or contracting $T_{\text{CNM}}$ along relevant modes. These support semantic similarity searches, clustering, and analogy finding \cite{lu_emergence_2019, obrien_machine_2024}.
\end{itemize}
This dual nature—a core computational tensor and derivable interpretable views—is fundamental to the \de{}. AI agents (Sec. \ref{subsec:agentic_synthesis_exploration}) primarily interact with and reason upon $T_{\text{CNM}}$ using operations from tensor algebra, geometric deep learning \cite{bronstein_geometric_2021}, or learned transformations \cite{kleyko_vector_2022, jr_computational_2018, frady_computing_2022}. For instance, identifying novel relationships might involve detecting unexpected non-zero entries in $T_{\text{CNM}}$ or applying tensor completion techniques. Synthesizing a new hypothesis (`KnowledgeArtifact`) could correspond to constructing a new subtensor pattern based on existing tensor components and desired properties. The insights or artifacts generated through these abstract computations are then translated back into the graph view or natural language for human comprehension and validation. This transformation of narrative scientific literature into a compact, computationally amenable tensor, from which various structured representations can be derived, forms the core of the \de{}'s approach to knowledge synthesis.


\subsection{Interoperability and FAIR Principles}

The structured, component-based nature of the CNM aligns fundamentally with the FAIR guiding principles\cite{wilkinson_fair_2016,jacobsen_fair_2020} for scientific data management and stewardship: Findable, Accessible, Interoperable, and Reusable. Knowledge components are \emph{Findable} through rich metadata and graph querying. They are \emph{Accessible} via standardized formats and potentially APIs built upon the CNM. \emph{Interoperability} is fostered by the use of controlled vocabularies (within the evolving template), potential links to standard ontologies, and the goal of using universal formalisms (like CT-inspired structures) for representation. Finally, the granular, verifiable, and context-rich nature of the components makes them highly \emph{Reusable} for subsequent analysis, synthesis, modeling, or integration into new research contexts. By design, the CNM aims to be a FAIR representation of scientific knowledge, moving beyond the limitations of isolated, opaque documents.

\section{Synthesizing the Field: Revealing the Knowledge Landscape}
\label{sec:synthesis}

Constructing the CNM involves more than just aggregating individual knowledge components; it requires synthesizing these elements to reveal the emergent structure, consensus, conflicts, and overall landscape of the research field. This synthesis process, facilitated by the pipeline component of the \de{}, transforms the collection of processed publications into a coherent, analyzable model.

\subsection{Aggregation and Principled Integration in Research Field Synthesis}

Following the structured distillation (Sec \ref{subsec:principles}), knowledge components from individual publications are integrated into the unified CNM graph. This process, central to field synthesis, moves beyond simple concatenation to intelligently combine information from diverse sources, resolve redundancies, align related concepts, and establish a consistent structure.

The aggregation begins by loading the structured knowledge components—representing node instances defined by the schema with their associated metadata and preliminary relationships—into the graph framework. These components form the initial "point cloud" derived from the literature corpus according to the evolved schema.

A primary challenge during this phase is \textbf{Entity Resolution and Alignment}. The same underlying scientific concept (TheoreticalConceptNode), method (MethodNode), material, etc., may be described differently across publications. The DE employs a multi-pronged strategy:
\begin{itemize}
 \item \textit{Semantic Alignment:} Utilizes derived vector embeddings associated with textual descriptions. Similarity searches (e.g., k-NN in embedding space \cite{kriegeskorte_representational_2008}) group potentially equivalent or related components, suggesting candidates for merging or linking.
 \item \textit{Structural Alignment:} Leverages the structured nature of the extraction. Components extracted using the same template probe (representing a consistent aspect or definition) are considered structurally aligned. Metadata comparison (e.g., units for ParameterNode, category for MechanismNode) provides further alignment evidence.
 \item \textit{Identifier Resolution:} Links components to standardized external identifiers (e.g., DOIs for PublicationNode, ontology terms, chemical identifiers) where available, providing unambiguous anchors.
\end{itemize}
Combining these evidence streams allows for robust identification and linking of components referring to the same entity, forming consolidated nodes within the CNM.

Once related components are aligned, \textbf{Principled Integration} determines how information from multiple sources is represented:
\begin{itemize}
 \item \textit{Quantitative Aggregation:} Multiple reported values for a parameter, ParameterNode can be aggregated statistically (mean, median, distribution fitting) to provide a synthesized estimate and uncertainty.
 \item \textit{Evidence Triangulation and Weighting:} Information prominence can be modulated based on source attributes (e.g., study type, sample size extracted via template) or network properties (e.g., citation impact of the PublicationNode, analogous to precision weighting in predictive processing architectures.
 \item \textit{Explicit Conflict Representation:} Contradictory findings are not ignored but explicitly captured, for instance by linking conflicting nodes or evidence via a dedicated KnowledgeGapNode, thereby highlighting areas requiring further investigation \cite{sejdinovic_hypothesis_2012}.
 \item \textit{Strict Provenance Tracking:} All synthesized information retains explicit links back to the originating source publications, ensuring full traceability.
 \item \textit{Formal Integration (Conceptual):} Principles from Category Theory, such as the colimit construction \cite{healy_episodic_2019}, offer a formal blueprint for merging structured information based on shared components while ensuring mathematical consistency, providing theoretical guidance for principled integration algorithms \cite{yuan_categorical_2023}.
\end{itemize}
This aggregation and integration process yields a unified, coherent knowledge graph where redundancies are minimized, related concepts interconnected, conflicts highlighted, and all assertions remain traceable to their evidentiary sources. This synthesized CNM forms the foundation for subsequent discovery-oriented analyses.

\subsection{Discovering Emergent Structures within the Synthesized Knowledge}

The \de{} leverages the synthesized CNM structure to automatically discover latent organizational patterns within the research field, reflecting its intrinsic structure rather than imposing rigid external classifications. Key techniques include:
\begin{itemize}
 \item \textit{Thematic Clustering:} Applying methods like BERTopic \cite{grootendorst_bertopic_2022} to semantic vector representations derived from node descriptions (\textit{e.g.}, mechanism definitions, publication abstracts) identifies dominant research themes and subfields, providing a semantic map of the knowledge landscape.
 \item \textit{Relational Network Analysis:} Employing algorithms from network science on the CNM graph structure identifies influential nodes (\textit{e.g.}, central concepts, foundational papers), bridge nodes connecting disparate thematic clusters, cohesive research communities, and recurring structural motifs (\textit{e.g.}, common experimental workflows involving sequences). Analyzing the temporal evolution of these structures, potentially using methods from scientometrics \cite{rong_40_2025} or complex systems \cite{buehler_self-organizing_2025}, reveals dynamic shifts in research focus and paradigm emergence, potentially uncovering dynamics analogous to self-organized criticality.
 \item \textit{Quantitative Landscape Synthesis:} Statistical analysis of aggregated quantitative data reveals typical value ranges, correlations between parameters and outcomes (e.g., linking material properties to system performance), identifies outliers requiring scrutiny, and establishes field-wide empirical distributions that contextualize individual findings.
\end{itemize}
These synthesis steps collectively transform the aggregated data into an interpretable model of the field's structure, dynamics, and quantitative landscape.

\subsection{The CNM as a Dynamic, Evolving Knowledge Structure}

Crucially, the CNM is conceptualized as a dynamic entity, distinct from static knowledge bases. Its evolution mirrors the ongoing scientific process \cite{buehler_agentic_2025-1, buehler_self-organizing_2025} through several mechanisms:
\begin{itemize}
 \item \textit{Continuous Integration:} New findings from ongoing publications are processed by the pipeline and integrated into the CNM graph, keeping the model current.
 \item \textit{Schema Adaptation:} The underlying extraction template is periodically refined via the self-consistent feedback loop (Fig. \ref{fig:template_evolution}), allowing the representational framework itself to adapt to evolving concepts and improve its fidelity.
 \item \textit{Dynamic Emergent Structures:} Thematic clusters, influential concepts, and network topology identified via synthesis are not fixed but evolve over time \cite{grootendorst_bertopic_2022, rong_40_2025}, providing insights into the meta-level dynamics of the scientific field.
\end{itemize}
This inherent dynamism, managed through appropriate versioning, ensures the CNM remains a relevant and adaptive foundation for ongoing scientific inquiry, transforming the knowledge base into a living model that reflects the research front.

\section{Navigating the Knowledge Landscape: Interaction and Discovery}
\label{sec:interaction}

The synthesized Conceptual Nexus Model (CNM) serves not only as a dynamic knowledge repository but, more importantly, as an interactive landscape for exploration and discovery. The \de{} provides mechanisms for users, both human researchers and AI agents \cite{wooldridge_1_1996}, to navigate this landscape, analyze information in context, and leverage the structure of the CNM to identify novel research opportunities.

\subsection{Contextual Placement and Analysis}

A fundamental capability enabled by the CNM is the contextualization of new information. When a researcher introduces a new query, a draft manuscript abstract, a set of experimental results, or references a recent publication, the DE processes this input using the same distillation and representation pipeline used to build the CNM itself. The resulting structured, vectorized components are then mapped onto the existing CNM landscape. Algorithms compute semantic proximity (using embeddings) and structural relationships (using graph connectivity or potentially VSA/CT formalisms) to identify the most relevant existing nodes, thematic clusters, and established relationships within the model \cite{kriegeskorte_representational_2008}. This mapping provides immediate context, indicating how the new information relates to the established body of knowledge: Does it align with existing consensus? Does it fall within a known thematic area? Does it address an identified gap? Does it potentially conflict with previous findings? This contextual placement transforms information retrieval into a deeper analytical process.

\subsection{Structure-Driven Exploration and Conceptual Traversal}

The \de{} facilitates modes of knowledge exploration fundamentally different from traditional linear document reading or keyword-based search paradigms. Users interact with the CNM by navigating its inherent relational structure, often visualized through interactive interfaces such as dynamic graph browsers or hyperlinked knowledge bases (conceptually similar to applications developed for specific databases \cite{koupil_unified_2022}). This structure-driven navigation allows researchers to follow diverse conceptual pathways through the synthesized landscape. For example, one can trace the implementation of specific scientific mechanisms (MechanismNode) across different experimental systems (SystemNode) via ImplementsMechanismEdge. Users can compare methodologies (MethodNode) applied to investigate related phenomena (PhenomenonBehaviorNode) by exploring neighboring nodes within identified thematic clusters (Sec. \ref{sec:synthesis}) or connected via specific relation types. Foundational concepts (TheoreticalConceptNode) or seminal publications (PublicationNode) can be identified by examining nodes with high centrality metrics derived from graph analysis. Critically, users can explore the frontiers of knowledge by investigating sparsely connected regions of the graph or by focusing on identified KnowledgeGapNode instances and their associated contextual nodes. The ability to systematically track the evolution of concepts or the usage of methods over time, by incorporating temporal metadata associated with nodes and edges, adds another dimension to exploration \cite{rong_40_2025}. This traversal, guided by the CNM's explicit structure and augmented by semantic proximity information, fosters a more holistic and nuanced understanding of a field's topology than is possible through fragmented document access \cite{shepard_second-order_1970}.


\subsection{Agent-Assisted Knowledge Synthesis and Generative Exploration on the CNM}
\label{subsec:agentic_synthesis_exploration}

The explicit, machine-readable, and richly interconnected structure of the \cnm{} (Sec. \ref{sec:cnm}) serves as an ideal operational environment for sophisticated AI agents, transforming the knowledge graph from a passive repository into a dynamic substrate for discovery \cite{wooldridge_1_1996}. Within the \de{} ecosystem, these agents are designed not merely to retrieve information, but to actively assist researchers in synthesizing existing knowledge and generating novel insights by leveraging the CNM's structural and semantic integrity.

AI agents within the \de{} can be tasked with several advanced functions that transcend traditional search and analysis capabilities:

\begin{itemize}
    \item \textbf{Automated Evidence Aggregation and Structured Synthesis:} Agents can systematically traverse the \cnm{}, following defined relational pathways (e.g., chains of `CAUSES` edges, or compositions of `IMPLEMENTS\_METHOD` and `PRODUCES\_OBSERVATION` links) to gather diverse yet interconnected knowledge artifacts pertaining to a specific scientific query or hypothesis. Beyond simple collection, these agents can employ summarization techniques or structured reasoning (akin to "thought graph" traversal \cite{besta_demystifying_2025}) to synthesize these components into coherent, evidence-backed narratives or structured summaries. For instance, an agent could construct a mechanistic explanation for a phenomenon by assembling a subgraph of relevant `MechanismNode`, supporting `PublicationNode`, and associated `Property/Parameter`, ensuring all links are justified by the CNM's provenance data \cite{sequeda_knowledge_2025}.

    \item \textbf{Complex Pattern Recognition and Inconsistencies Detection:} The CNM's graph structure is amenable to advanced analytical techniques. AI agents equipped with graph neural networks (GNNs) \cite{buehler_graph-aware_2025} or other pattern recognition algorithms can identify non-obvious recurring motifs (e.g., common methodological sequences, characteristic mechanistic patterns), complex correlations between disparate knowledge components, or deviations from established patterns (anomalies) that might signify emerging research fronts or inconsistencies requiring further investigation \cite{rong_40_2025}. Such capabilities enable a shift from hypothesis-driven queries to data-driven discovery of latent structures within the entire scientific field as represented by the CNM \cite{buehler_self-organizing_2025}.

    \item \textbf{Agent-Facilitated Generation of Novel Knowledge Artifacts:} A core innovation of the \de{} lies in its support for AI agents to assist in the creation of new scientific knowledge (see Sec. \ref{sec:discovery} for hypothesis generation). This moves beyond mere exploration to active construction. For example:
        \begin{itemize}
            \item \textit{Analogical Transfer:} An "Analogy Agent," leveraging both semantic similarity from vector embeddings and structural isomorphism identified within the CNM (potentially inspired by category-theoretic notions of functors \cite{phillips_what_2022, yuan_categorical_2023}), could propose adapting a successful `Process/Method` from one `Entity/System` context to another, generating a novel experimental design artifact \cite{lu_emergence_2019, obrien_machine_2024}.
            \item \textit{Compositional Design:} Users, interacting with agents on a conceptual workbench (further detailed in design-focused documentation for the \de{} platform), can combine existing CNM components (e.g., `MechanismNode`s, `MaterialNode`s) in novel ways. AI agents would provide real-time feedback on the structural validity (adherence to the Universal Concept Scheme, Appendix \ref{app:schema}) and potential plausibility (based on known constraints or similar existing patterns in the CNM) of these new compositions, assisting in the assembly of new `KnowledgeArtifactNode` instances representing complex hypotheses or system designs \cite{he_two-way_2024}.
            \item \textit{Targeted Gap Filling:} When a \texttt{KnowledgeGapNode} is identified, AI agents can be tasked to search the CNM for components that, if combined or modified, could plausibly address that gap, effectively proposing research directions \cite{buehler_agentic_2025-1}.
        \end{itemize}
\end{itemize}

The operational premise is that these AI agents are tools whose reasoning processes (especially those involving graph traversal or component assembly) can be made transparent by virtue of operating on the explicit and verifiable structure of the CNM \cite{manzoor_expanding_2023}. Human researchers interact with these agents, guiding their exploration, refining their suggestions, and ultimately validating the new knowledge artifacts generated. This symbiotic human-AI interaction, grounded in a shared, structured understanding of the scientific domain (the CNM), is what the \de{} framework aims to foster, moving scientific inquiry towards a more synthesized, computationally augmented, and generative paradigm \cite{tsividis_human-level_2021}. The specifics of user interaction with these agents, including user-configurable agent behaviors and collaborative interfaces, are elaborated in complementary work focusing on the \de{} platform design.

\section{Automated Gap Analysis and Hypothesis Generation}
\label{sec:discovery}

Beyond facilitating understanding of existing knowledge, a primary goal of the \de{} is to actively catalyze the discovery of new knowledge by supporting the iterative cycle of scientific inquiry. This cycle can be viewed as a process of refining models of the world based on evidence – starting with prior understanding, gathering informative data, updating beliefs (models), and using that updated understanding to generate new hypotheses and guide further investigation \cite{HowsonUrbach2006}. By leveraging the synthesized structure of the CNM, the platform aims to systematically identify promising research opportunities (areas where belief updating is most needed) and generating scientifically verifiable hypotheses grounded in the current state of knowledge.

\subsection{Systematic Identification of Knowledge Gaps}

The explicit representation of what is known within the CNM simultaneously highlights what is unknown or poorly understood. The DE employs algorithmic techniques to automatically detect various types of knowledge gaps:

\emph{Component Completeness Gaps} arise when specific types of information, defined as essential by the template (e.g., specific parameters, control mechanisms, quantitative performance metrics), are consistently missing for certain classes of systems or concepts represented in the CNM.

\emph{Structural Holes} refer to regions within the knowledge graph where expected connections between related concepts, thematic clusters, or steps in a process are sparse or absent \cite{buehler_agentic_2025-1}. Identifying these "missing links" points towards unexplored relationships or required intermediate steps.

\emph{Inconsistency Clusters} emerge when the CNM reveals groups of structurally similar components (e.g., experiments using comparable methodologies) that report contradictory findings or support conflicting theoretical interpretations. These represent areas where the current knowledge is contested and requires resolution through further investigation \cite{sejdinovic_hypothesis_2012}.

\emph{Predictive Gaps} can be identified by leveraging the structural regularities or theoretical principles captured within the CNM. For example, based on established patterns or analogies identified through graph analysis or CT-inspired reasoning \cite{buehler_graph-aware_2025, phillips_what_2022}, the system might predict the existence of a certain mechanism or relationship that has not yet been reported in the literature, thus identifying a specific target for empirical confirmation.

By systematically surfacing these diverse types of gaps, the DE provides researchers with a data-driven map of the scientific frontier, highlighting areas where new research is most needed or likely to be impactful.

\subsection{Principled Hypothesis Generation}

The identification of gaps and structural patterns within the CNM serves as a foundation for generating novel, testable hypotheses. The DE aims to move beyond simple correlation-finding towards true abductive inference: generating hypotheses grounded in mechanistic understanding and structural analogy:

\emph{Bridging Gaps with Existing Knowledge:} Hypotheses can be formulated to directly address identified gaps. For instance, if a crucial parameter is missing for a class of systems, the hypothesis might involve designing an experiment to measure it. If a structural hole exists between two related concepts, the hypothesis might propose a specific mechanism linking them. Specific informative experiments can be proposed, either heuristically (e.g. choosing experimental parameters not yet empirically investigated), or more formally (e.g. with an explicit information gain term). 

\emph{Exploiting Structural Analogies:} By identifying non-obvious structural similarities between components from different thematic clusters or even different disciplines represented within the CNM (e.g., using graph isomorphism techniques \cite{buehler_graph-aware_2025} or CT functors \cite{phillips_what_2022}), the DE can generate hypotheses based on analogy. For example, it might suggest that a successful control strategy \cite{bechtel_grounding_2021}) from one system type could be adapted to address a limitation in a structurally analogous system from a different cluster \cite{lu_emergence_2019, obrien_machine_2024}. This structured analogy finding is potentially more powerful than relying solely on semantic similarity.

\emph{Resolving Conflicts through Synthesis:} Identified inconsistencies can spur hypotheses that propose novel mechanisms, theories, or experimental conditions capable of reconciling the conflicting observations represented within the CNM.

Noteworthy, hypotheses generated via these mechanisms are not unconstrained speculations but are directly derived from, and justifiable by, the synthesized structure of existing scientific knowledge captured within the CNM. This grounding aims to increase the relevance and potential impact of computationally generated research suggestions, positioning the DE as a genuine tool for augmenting scientific creativity and strategic research planning.

\section{Implementation, Validation, and Future Vision}
\label{sec:validation_discussion}

The conceptual framework of the \de{} requires practical implementation and validation to demonstrate its feasibility and utility. This section briefly outlines the implementation approach, summarizes validation strategies, and discusses the broader vision for the DE as an evolving infrastructure for scientific knowledge.

\subsection{Case studies}

\section{Case Studies: Applying the \de{} Framework}
\label{sec:case_studies}

To illustrate the capabilities and potential impact of the \de{} framework, we present two distinct case studies. The first demonstrates the \de{}'s application to synthesize a nascent scientific field and collaboratively construct a forward-looking perspective. The second highlights how the principles underlying the \de{} were meta-applied to assist in the conceptual design of an interactive platform for knowledge exploration itself.

\subsection{Case Study 1: Synthesizing and Shaping the Field of Intelligent Soft Matter}
\label{subsec:casestudy_softmatter}

The emerging field of "Intelligent Soft Matter" lies at the dynamic intersection of materials science, physics, chemistry, biology, and cognitive science. It aims to create materials with life-like cognitive capabilities such as perception, learning, memory, and adaptive decision-making, moving beyond traditional passive or simply responsive materials \cite{baulin_intelligent_2025}. Given its interdisciplinary nature and rapid evolution, this field presents an ideal testbed for the \de{}'s ability to synthesize knowledge and identify a consensual research trajectory.

An initial corpus of key publications relevant to intelligent soft matter was processed using the \de{} methodology (Sec. \ref{sec:cnm}). This involved:
\begin{enumerate}
    \item \textbf{Initial Template Design and Distillation:} A preliminary extraction template, based on the Universal Concept Schema (Appendix \ref{app:schema}) and tailored with probes specific to soft matter and embodied intelligence, was used to guide LLMs in distilling knowledge artifacts from the selected papers.
    \item \textbf{Iterative Template Refinement and CNM Construction:} A diverse group of researchers active in fields contributing to intelligent soft matter engaged with the initial distilled components and the template itself. Through a series of iterative feedback cycles, managed within a collaborative environment built on \de{} principles, the template was refined. Specialized AI agents, trained on subsets of the evolving \cnm{} pertaining to specific material classes (e.g., hydrogels, LCEs) or mechanisms (e.g., self-organization, memory encoding), assisted in identifying ambiguities in the template or inconsistencies in the extracted data. For example, an LLM agent trained on materials papers might flag that the template inadequately captured parameters related to swelling kinetics critical for adaptive responses, prompting a template revision.
    \item \textbf{CNM Synthesis and Gap Analysis:} As the template converged towards a consensus representation reflecting the shared understanding of the involved experts and the literature corpus, a \cnm{} for intelligent soft matter was constructed. AI agents then analyzed this CNM to identify overarching themes (using methods similar to those in Sec. \ref{sec:synthesis} B), key conceptual hubs, and critical knowledge gaps (e.g., lack of robust methods for quantifying material 'learning').
    \item \textbf{Collaborative Perspective Generation:} The synthesized CNM, along with the identified themes and gaps, served as the structured foundation for a collaborative effort involving domain experts and AI. This led to the generation of a forward-looking perspective on the field, outlining key challenges, promising research directions, and a conceptual roadmap for realizing materials with true intelligent behavior \cite{baulin_intelligent_2025}. The AI agents assisted in drafting sections based on specific CNM subgraphs, ensuring claims were grounded in the synthesized evidence, and helping to maintain consistency across contributions from multiple human authors.
\end{enumerate}

This case study demonstrates how the \de{}, through its iterative, AI-assisted, and collaborative approach to template refinement and knowledge synthesis, can facilitate the consolidation of an emerging scientific field. It transformed an array of individual publications into a structured \cnm{} that not only represents existing knowledge but also serves as a generative substrate for defining the field's future trajectory. The resulting perspective \cite{baulin_intelligent_2025} showcases a collaboratively constructed understanding, richer and more systematically grounded than what might emerge from individual expert reviews alone.

\subsection{Case Study 2: AI-Assisted Design of the DE Platform Interaction Framework}
\label{subsec:casestudy_platform_design}

The principles of structured knowledge synthesis and agent-assisted generation inherent in the DE framework were meta-applied to inform the conceptual design of the DE platform itself—the interactive environment for human-AI collaboration detailed in complementary work \cite{lumiruusu_resnei_2025}.

\begin{figure}[ht]
    \includegraphics[scale=0.25]{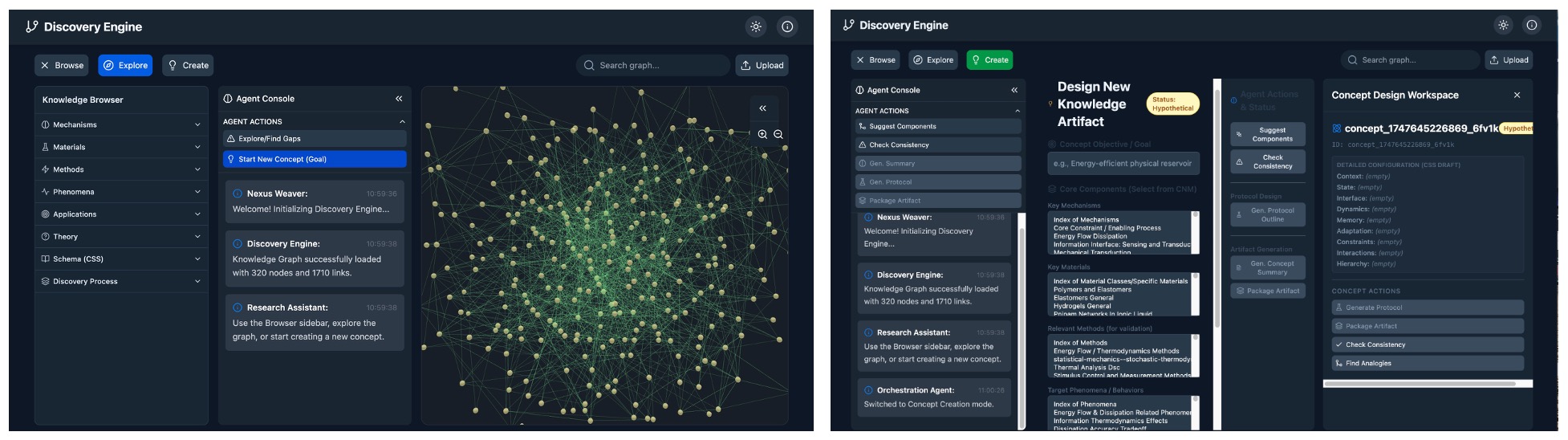}
    \caption{User Interface (UI) for the DE platform, fully generated through an AI-assisted design process. The core modules (e.g., graph visualization, knowledge browser, agent interaction panel, hypothesis workbench) and their relationships were initially outlined by DE (with Gemini Pro 2.5) synthesizing best practices from HCI and KG interaction literature, then refined by a human.}
    \label{fig:platform_ui_concept}
\end{figure}

\begin{enumerate}
    \item \textbf{Distillation of HCI and KG Interaction Literature:} A corpus of relevant research papers focusing on human-computer interaction (HCI) for complex data, knowledge graph visualization, explainable AI (XAI), and collaborative systems (including those cited throughout this manuscript, e.g., \cite{li_knowledge_2024, sarrafzadeh_hierarchical_2020, zhang_patterns_2021, he_two-way_2024, sequeda_knowledge_2025, meier_classify_2023, rahman_knowledge_2024}) was processed using an early version of the \de{} distillation pipeline. The extraction template focused on identifying core interaction principles, user interface patterns, described user challenges, proposed AI assistance roles, and evaluation methodologies.
    \item \textbf{Synthesis of Design Principles and Feature Requirements:} The distilled components were synthesized into a CNM focused on "KG Interaction Design." AI agents were then used to analyze this specialized CNM.
        \begin{itemize}
            \item \textit{Pattern Identification:} Agents identified recurring successful interaction patterns (e.g., multi-modal views, focus+context, provenance tracking) and common usability challenges (e.g., query complexity, information overload).
            \item \textit{Principle Extraction:} Based on this analysis, core design principles for the DE platform were formulated (e.g., "Support Multi-Modal Exploration," "Ensure Verifiability and Provenance," "Facilitate Human-AI Co-Creation," "Manage Cognitive Load").
        \end{itemize}
    \item \textbf{AI-Assisted Generation of Platform Concepts and UI Mockups:} Working from these principles and the synthesized interaction patterns, designers collaborated with generative AI tools (LLMs prompted with the distilled requirements and design principles). This collaboration yielded conceptual designs for key platform modules, such as the multi-modal exploration interface, the "Knowledge Card" summaries, and the "Hypothesis Workbench." AI was used to generate initial textual descriptions of module functionalities and even draft visual concepts for the user interface (an example conceptual UI schema generated through this process is shown in Fig. \ref{fig:platform_ui_concept}).
    \item \textbf{Iterative Refinement:} These AI-generated concepts were then iteratively refined by human designers and HCI experts, ensuring alignment with user needs and established HCI best practices. This mirrored the `process.md` workflow where AI provides initial drafts that humans then curate and enhance.
\end{enumerate}

The Discovery Engine (DE) frontend is implemented as an interactive React/TypeScript application (see \url{https://github.com/ActiveInferenceInstitute/Research-Discovery-Engine}). It processes Markdown (.md) files from a KG/ directory via a cnmBuilder.ts utility to construct a client-side graph representation—the Conceptual Nexus Model (CNM). Key UI components include: an AgentConsole for simulated agent interactions and workflow guidance; a GraphVisualization module using 3d-force-graph for exploring the CNM; a KnowledgeBrowserSidebar for navigating the content hierarchy of the source .md files; and a NodeView for displaying detailed content of selected knowledge graph nodes. A ConceptDesigner module allows users to define new hypothetical SystemNode configurations based on components from the CNM, with future integration points for AI-driven suggestions and validation. The current system demonstrates the client-side parsing, graph construction, and interactive visualization pipeline, with persistent knowledge artifacts intended to be generated from the design process. While advanced LLM-driven extraction and algorithmic synthesis (like topic modeling) are conceptualized as backend processes in the DE system, the frontend focuses on structuring and presenting the user-guided discovery workflow.

\subsection{The CNM Structure as an Evolution of Scientific Communication}

The DE methodology offers a potential pathway to address the systemic challenges facing scientific communication and progress in an era of exponential information growth. By shifting the focus from individual documents to synthesized, verifiable knowledge components organized within a dynamic CNM, it aims to create a more robust, navigable, and computationally accessible representation of collective scientific understanding.

This structured synthesis provides crucial context that is often lost in fragmented literature. It enables researchers to quickly situate new findings, objectively assess consensus and conflict, and identify the implicit structure and trajectory of their field. The ability to compare components across studies based on a consistent schema facilitates deeper understanding than narrative summaries alone can provide.

Its gap analysis and hypothesis generation features directly support the crucial steps of identifying where new evidence is most needed and proposing informative experiments, aligning with formal frameworks of rational inquiry like Bayesian reasoning and potentially Active Inference\cite{HowsonUrbach2006,Friston2017Active}.

The challenges of information overload, reproducibility concerns \cite{baker_1500_2016}, the proliferation of potentially unreliable AI-generated content, and potentially declining research disruptiveness \cite{park_papers_2023} necessitate a fundamental rethinking of how scientific knowledge is structured, disseminated, and utilized. The traditional narrative publication, often distributed as a static PDF, faces limitations in this new environment.

The DE methodology and the resulting CNM structure offer a potential evolution, acting as a complementary layer built upon the primary literature. The CNM's structure directly addresses several key challenges:

Its synthesized nature combats \emph{information overload} by providing structured overviews. Its focus on verifiable components linked to source evidence acts as a filter against low-quality or fabricated content, including potential \emph{LLM spam}. The explicit capture of methods and parameters enhances \emph{reproducibility} assessment. The graph structure facilitates deep \emph{synthesis} beyond individual document summaries. Crucially, the CNM is designed to be inherently machine-readable and structured, aligning perfectly with the principles of \emph{FAIR data} (Findable, Accessible, Interoperable, Reusable) \cite{wilkinson_fair_2016,jacobsen_fair_2020}. This computational accessibility is vital for leveraging AI tools effectively and responsibly within the scientific process.

In future similar machine-readable structures can replace narrative papers entirely, however, for past knowledge reflected in numerous PDFs the CNM could become the primary interface for computational analysis, synthesis, and discovery, providing a dynamic, validated, and navigable map derived from the underlying literature. This represents a shift towards a more interconnected, computable, and potentially more efficient scientific knowledge ecosystem.

\subsection{Limitations and Future Directions}

Significant challenges remain. The DE's effectiveness is contingent on the continued advancement of LLMs for nuanced scientific text understanding \cite{rogers_primer_2021, lu_fine-tuning_2025} and the careful design and governance of the evolving template. Representing the full complexity of scientific knowledge, including uncertainty, causality, and temporal dynamics, within computationally tractable frameworks requires ongoing research, potentially integrating probabilistic methods \cite{sejdinovic_hypothesis_2012, sejdinovic_equivalence_2013}, Category Theory \cite{yuan_categorical_2023, pan_token_2024, hefford_categories_2020, phillips_what_2022, koupil_unified_2022, purvine_category_2016, christino_theoretical_2023}, or VSA techniques \cite{kleyko_vector_2022, jr_computational_2018}. However, scalability to disciplines with millions of publications poses substantial engineering hurdles \cite{buehler_agentic_2025-1}.

Future work will focus on several key areas: enriching the CNM representation to capture deeper semantic and causal relationships; enhancing the data processing pipeline for greater robustness and efficiency; developing more powerful AI agent capabilities for synthesis, gap analysis, and hypothesis generation, perhaps drawing inspiration from Active Inference \cite{ciaunica_nested_2023, tsividis_human-level_2021} or multi-agent systems \cite{wooldridge_1_1996, besta_demystifying_2025, he_two-way_2024, zhao_agentigraph_2024, manzoor_expanding_2023}; and creating intuitive, interactive interfaces for human researchers to collaborate with the \de{} \cite{li_preliminary_2024, meier_classify_2023, rahman_knowledge_2024, sarrafzadeh_knowledge_2016, skjaeveland_ecosystem_2024, zhang_patterns_2021}. A long-term vision includes integrating the DE with automated experimental platforms or simulation engines, creating a closed loop from knowledge synthesis to hypothesis generation to empirical testing, thereby truly accelerating the cycle of scientific discovery.

\section{Conclusion}

The \de{} framework addresses the urgent need for more effective mechanisms to synthesize and navigate the ever-expanding body of scientific knowledge. Our methodology moves beyond document-centric approaches by first distilling publications into verifiable, structured knowledge artifacts using LLMs guided by adaptive templates. The core innovation lies in encoding this rich, heterogeneous information into a unified Conceptual Nexus Tensor ($T_{\text{CNM}}$). This tensor serves as a compact, machine-operable "World Model" of a scientific domain, capturing complex interdependencies within its high-dimensional structure.

From this central tensorial representation, human-interpretable views, such as explicit knowledge graphs (the CNM graph) and semantic vector spaces, can be dynamically generated, enabling researchers to explore the synthesized landscape. More profoundly, $T_{\text{CNM}}$ provides a substrate for specialized AI agents to perform complex reasoning, identify structural patterns, uncover latent analogies, and pinpoint knowledge gaps through abstract mathematical operations. These agents then collaborate with human researchers, assisting in the construction of novel knowledge artifacts—hypotheses, experimental designs, or theoretical models—that are grounded in the synthesized evidence contained within the tensor.

While the full realization of dynamic tensorial knowledge bases presents significant research and engineering challenges, the \de{} framework outlines a principled pathway. By transforming disparate scientific narratives into a structured, computable tensor, and by enabling AI-assisted interaction with this representation, the \de{} aims to create a generative ecosystem for scientific discovery. 

By revealing the topology of knowledge such as clusters, connections, and, significantly, its gaps and inconsistencies, the \de{} aims to transform discovery from a process often reliant on serendipity into a more systematic exploration guided by the structure of what is known and unknown in a more informed way. It provides a framework for identifying non-obvious connections, formulating targeted questions, and generating novel hypotheses grounded in the synthesized evidence. While challenges remain in implementation, scalability, and validation, the \de{} represents a powerful vision for augmenting human intellect, fostering deeper understanding, and accelerating the pace of scientific breakthroughs in an increasingly complex world. It endeavors to build not just better search engines, but genuine engines of discovery, potentially reshaping the future of scientific communication and collaboration.

\appendix
\section{General Structure of a Distillation Template}
\label{app:template}
This template is designed to be a structured set of instructions and placeholders used by an AI agent (specifically a Large Language Model - LLM) to systematically extract key information from a single scientific publication. The goal is to transform the unstructured (or semi-structured) text of a paper into a collection of discrete, well-defined "knowledge components" and their relationships. These extracted components are then used to populate your Conceptual Nexus Model (CNM), which is built according to the Universal Concept Schema (UCS).

The template is divided into major modules (\textit{e.g.}, M0: Meta-Information, M1: Core Claims, M2: Entities \& Systems).
This organization helps break down the complex task of analyzing a whole paper into more manageable sub-tasks for the LLM.
It also ensures that different categories of information are systematically considered. Within each module, there are specific "probes" or fields. These are direct instructions or questions to the LLM.

\section{A Universal Concept Schema for General Scientific Knowledge Representation and Synthesis}
\label{app:schema}

\begin{figure}[tbh]
 \centering
 \includegraphics[width=0.9\textwidth]{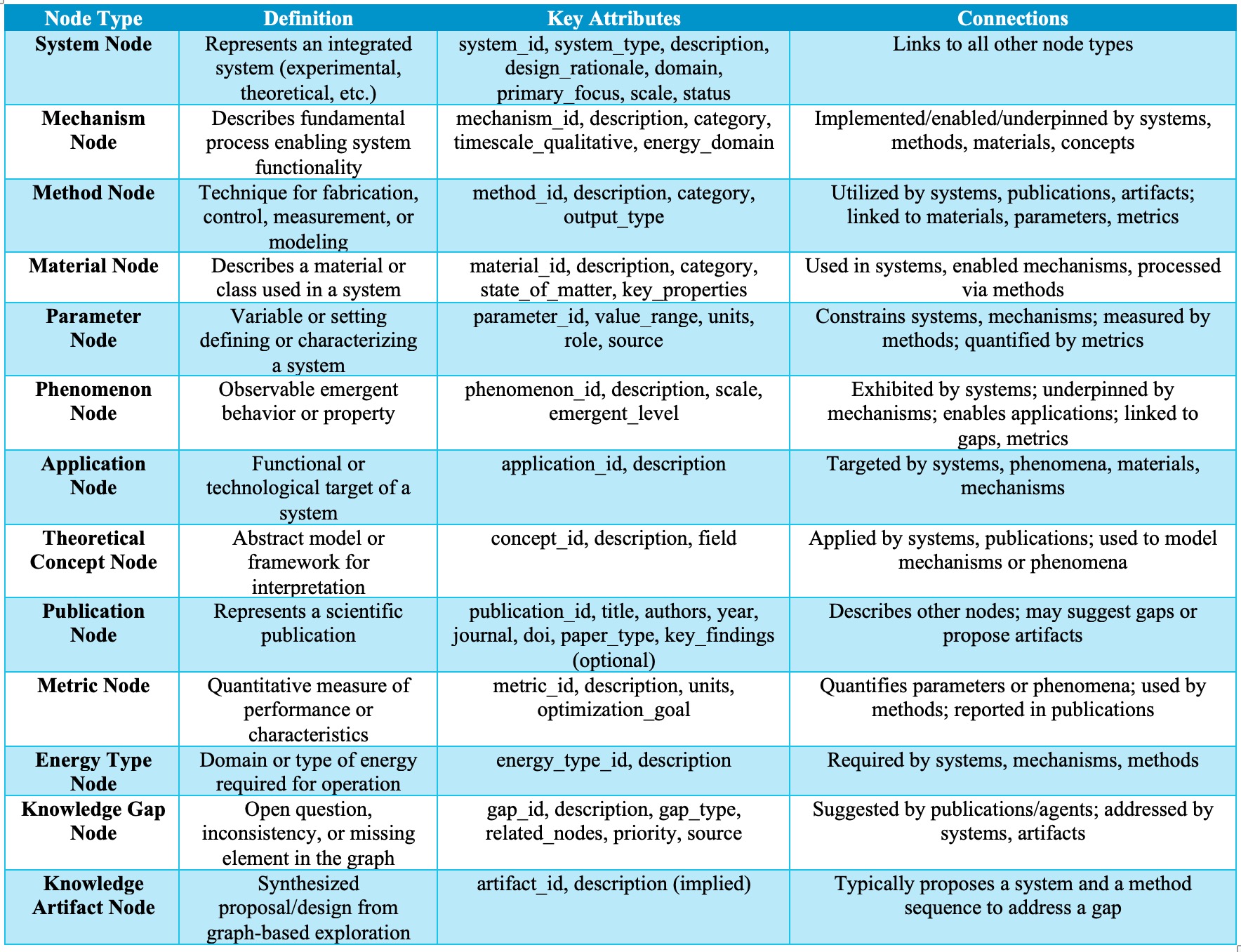}
 \caption{CNM Nodes, their definitions and key attributes}
 \label{fig:nodes}
\end{figure}

The primary objective is to transform disparate scientific information into a dynamic, interconnected network of "knowledge components" that align with FAIR principles, that can represent the analogy of a "world model" for scientific discoveries. This network, far from being a static archive, is designed as an active substrate for computational reasoning, enabling the identification of knowledge gaps and, crucially, facilitating the generation of novel "Knowledge Artifacts". These artifacts may represent new hypotheses, proposed experimental designs to test specific theories, innovative theoretical models, or conceptual blueprints for new systems or methods. Its instances that populate the CNM.

\begin{figure}[tbh]
 \centering
 \includegraphics[width=0.9\textwidth]{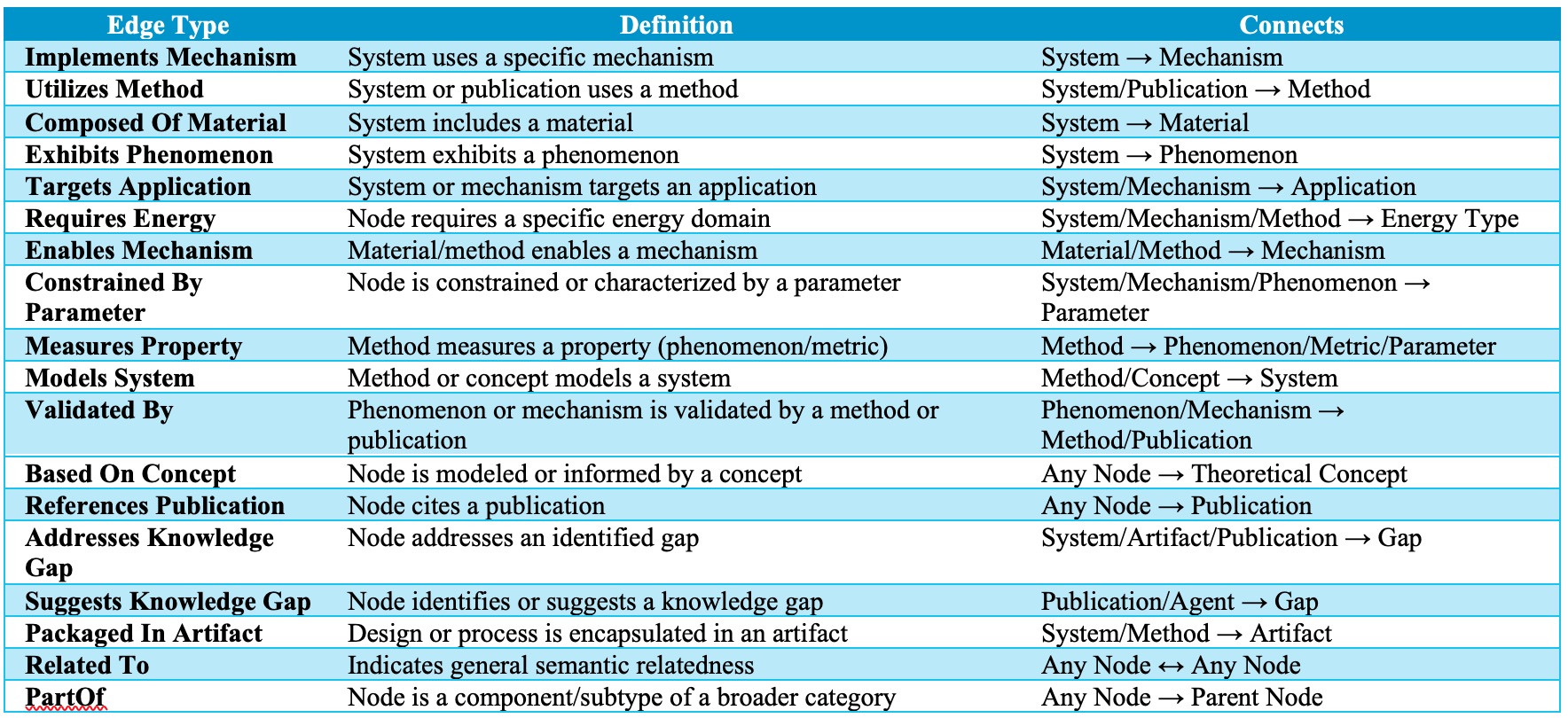}
 \caption{CNM Edges, their definitions and key attributes}
 \label{fig:edges}
\end{figure}

\subsection*{I. Core Node Archetypes (Conceptual Classes)}

We propose a minimal set of fundamental node archetypes to represent the universal building blocks of scientific knowledge. These are conceptual classes; specific instances derived from the literature or generative processes would populate the actual knowledge graph.

\subsection*{II. Core Relationship Archetypes (Conceptual Edge Types)}

These define the fundamental ways in which the node archetypes connect, forming the relational fabric of the scientific knowledge graph.
\begin{figure}[tbh]
 \centering
 \includegraphics[width=0.9\textwidth]{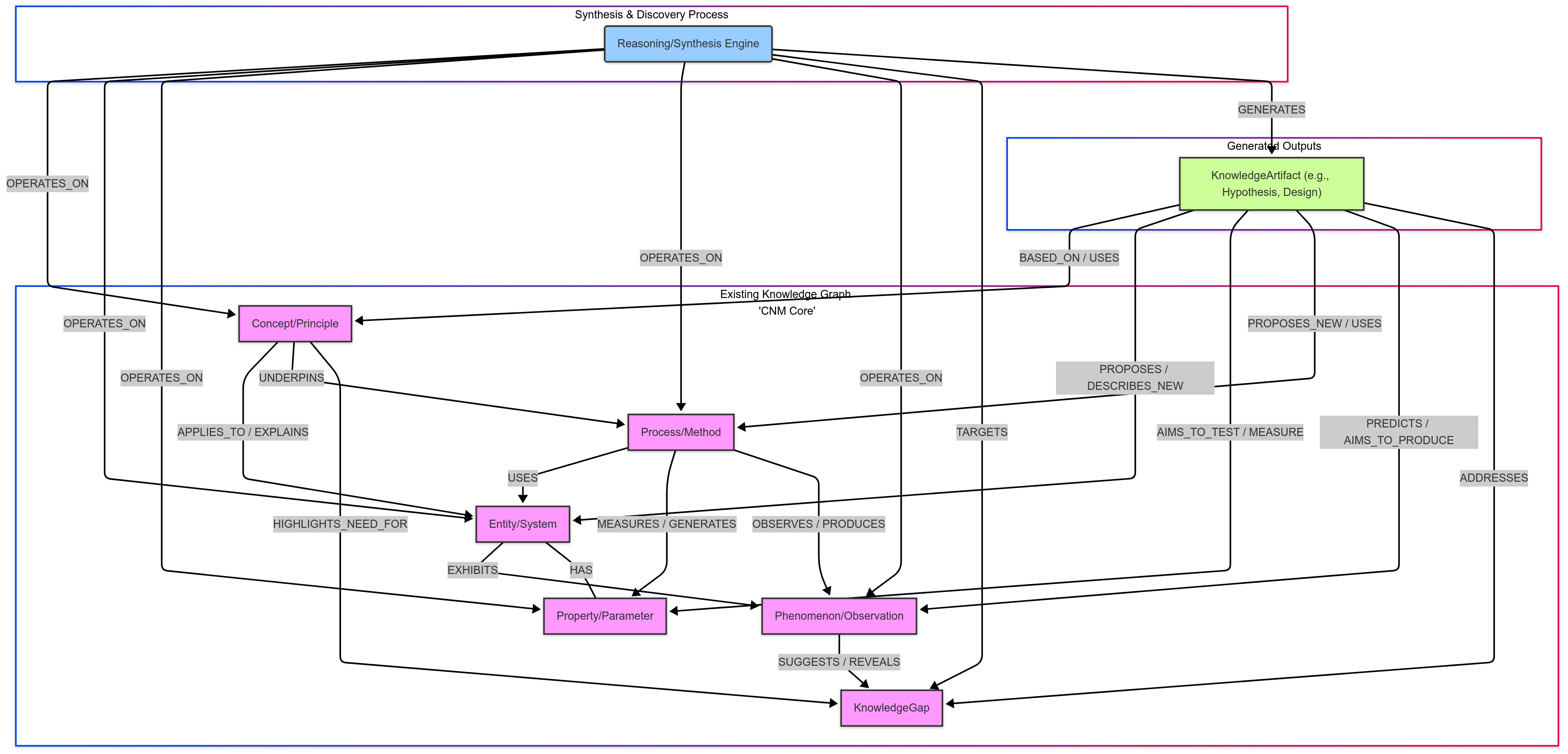} 
 \caption{A conceptual representation of the core knowledge graph structure emphasizing the generation of Knowledge Artifacts. Existing knowledge components (Concepts, Entities, etc.) form a network that is analyzed by a Reasoning/Synthesis Engine. This engine, by identifying and targeting Knowledge Gaps, generates new Knowledge Artifacts (e.g., hypotheses, designs, models), which are themselves composed of, based on, or aim to validate existing and new components.}
 \label{fig:css_concept_graph}
\end{figure}

\subsection*{III. Interaction between Node Archetypes}

The interaction between these archetypes can be visualized as a dynamic graph where the core process involves reasoning over the existing knowledge landscape (comprising Concepts, Entities, Properties, Methods, and Observations) to identify Knowledge Gaps and, critically, to synthesize novel Knowledge Artifacts designed to address these gaps or explore new frontiers (Fig. \ref{fig:css_concept_graph}).


%

\end{document}